\documentstyle[floats,aps,prb,epsf,eqsecnum]{revtex}
\newif\ifbig

\bigtrue
\bigfalse

\ifbig
        
\fi
\begin{document}
\draft
\date{\today}


\def\varepsilon{\epsilon}

\def\lambdab{{\bar \lambda}}
\def\mub{{\bar \mu}}
\def\Yb{{\bar Y}}
\def\cb{{\bar c}}
\def\kb{k_{\rm B}}

\def\fd{{\rm (fd)}}
\def\sur{{\rm (s)}}
\def\dis{{\rm (d)}}
\def\imp{{\rm (i)}}
\def\ext{{\rm (e)}}
\def\c{{\rm (c)}}
\def\cimp{{\rm (i)}}
\def\cfd{{\rm (fd)}}

\def\Dsur{\Delta_\sur}
\def\Dimp{\Delta_\imp}
\def\Dfd{\Delta_\fd}

\def\gammab{{\bar \gamma}}
\def\sigmab{{\bar \sigma}}
\def\dgamma{{\bar{\delta\gamma}}}
\def\bfd{b_\fd}

\def\r{{\bf r}}
\def\b{{\bf b}}
\def\u{{\bf u}}
\def\ex{{{\bf e}_x}} 
\def\ez{{{\bf e}_z}} 
\newcommand{\bnabla}{{\bbox \nabla}}

\newcommand{\inta}[1]{\int_{-a}^a \! \mbox{d} #1\,}
\newcommand{\into}[1]{\int_{-1}^1 \! \mbox{d} #1\,}
\newcommand{\intt}[1]{\int \! \mbox{d}^2 {\bf #1}\,}
\newcommand{\intinta}[2]{\int_{-a}^a \int _{-a}^a \! 
\mbox{d} #1 \mbox{d} #2 \,}
\newcommand{\num}[2]{#1 \!\cdot\! 10^{#2}}
\newcommand{\enum}[1]{10^{#1}}
\newcommand{\dav}[1]
{
\left\langle #1 \right\rangle
}

\newcommand{\ocite}[1]{Ref.\ \onlinecite{#1}}

\ifbig
{}
\else
\twocolumn[\hsize\textwidth\columnwidth\hsize\csname@twocolumnfalse%
\endcsname
\fi
\title{A New Criterion for Crack Formation in Disordered Materials}
\author{Peter F. Arndt and Thomas Nattermann}
\address{Institut f\"ur Theoretische Physik,
        Universit\"at zu K\"oln, 
        Z\"ulpicher Str. 77, 
        D-50937 K\"oln, Germany}
\maketitle
\begin{abstract}
Crack formation  is conventionally described as a nucleation phenomenon despite
the fact that the temperatures necessary to overcome the nucleation barrier are
far too high. In this paper we consider the possibility that cracks are created
due to the presence of frozen disorder (e.g. heterogeneities or frozen
dislocations). In particular we calculate the probability for the occurrence of
a critical crack in a quasi two-dimensional disordered elastic system. It turns
out that this probability takes the form of an Arrhenius law (as for thermal
nucleation) but with the temperature $T$ replaced by an effective {\it disorder
temperature} $T_{\rm eff}$ which depends on the strength of the disorder. The
extension of these results to $d=3$ dimensions is briefly discussed.
\end{abstract}
\ifbig
{}
\else
]
\fi


\section{Introduction}
Cracks are one of the most important sources for the failure of solids
\cite{FiMa}.  Despite continuous efforts for more than a century a full
understanding of fracture has not yet been reached \cite{Lawn,Freund,HR}.  A
simple but very appealing picture for the occurrence of cracks goes back to
Griffith \cite{Griff}. Griffith describes the emergence of cracks as a
nucleation phenomenon: To open a crack in a thin plate, atomic bonds have to be
broken and two new surfaces have to be created.  For a crack of linear size
$a$, this costs an energy of the order $a$.  Simultaneously, the potential
energy of the plate under external load is reduced due to the stress relaxation
around the crack.  This decreases the energy by an amount of order $a^2$. Thus,
the total crack energy as a function of $a$ increases for small $a$ linearly
and reaches a maximum at $a=a_c$ before it decreases quadratically.  Cracks of
length $a<a_c$ are stable whereas those with $a>a_c$ are unstable.  However,
contrary to conventional nucleation phenomena the typical energy barriers for
crack propagation in a perfect solid under realistic stresses are much too high
to be overcome by thermal fluctuations.  Instead, the pre-existence of
micro-cracks (or pre-weakened bonds) on scales $a \lesssim a_c$ is usually
tacitly assumed.  These will then grow under external load.  It seems to be
reasonable to consider micro-cracks as well as other heterogeneities in the
material as a kind of frozen disorder only amenable to a statistical treatment.

The {\em propagation} of supercritical cracks in an inhomogeneous material has
been the subject of a number of articles \cite{Fis,Ball} which have attempted
to explain the roughness of crack fronts found experimentally \cite{EBou}.
Unfortunately a convincing explanation of the experimental data is still
lacking.  The other aspect, the {\em occurrence} of a critical crack in the
first place has been, to the best of our knowledge, not yet considered.  It is
the aim of the present paper to address this point by calculating the
probability of the occurrence of a critical crack in systems which includes
various types of disorder. In particular, we will consider randomness in the
atomic bond strength as well as randomly distributed impurities (or other kinds
of heterogeneities) and frozen dislocations.  It should however be mentioned,
that our considerations are not restricted to crystalline material. The main
ingredients of our theory is isotropic elasticity, which also applies to
amorphous materials, supplemented by randomly distributed disorder. The latter
can also include mesoscopic heterogeneities and microcracks which occur during
fabrication.  Although the various sources of disorder conceivable may differ
considerably in their local properties, the most important aspect from the
statistical point of view (which is adopted in this paper) is the spatial decay
of the stress fields they create. It is in this sense that the three types of
disorder considered below are generic.

The main body of the paper is related to crack creation in a thin plate of
infinite extension. Some results can however be easily extended to bulk
materials.  The detailed investigation of cracks in slabs of finite width as
well as those in three dimensional systems will be presented in forthcoming
publications \cite{AN}.

\section{Cracks in an infinite two-dimensional sample}
\label{sec:2dim_sample}
In this paper we consider a single planar crack extended parallel to the
$x$-direction in a 2-dimensional elastically isotropic system of infinite
extension.  The 2-dimensional situation can be realized e.g. by a thin plate of
width $h$ in the so-called plane stress configuration \cite{Filon}.  The
Lam{\'e} coefficients $\lambdab$ and $\mub$ of the 2-dimensional system are
then related to the Lam{\'e} coefficients $\lambda$ and $\mu$ of the bulk by
        \begin{math}
        \lambdab=2 \lambda \mu h/(\lambda + 2\mu )
        ,\,
        \mub=\mu h
        \end{math}.
The coordinates of the crack are given by
        \begin{equation}
        -a\le x\le a\,,\quad y\equiv 0\,.
        \label{eq:2.1}
        \end{equation}
Such planar crack appears for instance in experiments with preweakened bonds
\cite{DeScMa}.

In two dimensions only mode I and mode II cracks occur characterized by the
only non-zero components $\sigmab_{yy}^\ext=\sigmab_\ext$ or
$\sigmab_{xy}^\ext=\sigmab_{yx}^\ext=\sigmab_\ext$ respectively, of
the applied stress $\sigmab_{ij}^\ext$.  In the further treatment we will apply
the dislocation theory of cracks \cite{We}: The crack will be described by
virtual lattice planes filling its interior such that there is no
free crack surface.  The lattice planes terminate in crack dislocations with
Burgers vector ${\bf b}^\c$. The whole crack is then given by a collection of
dislocations (and antidislocations) ${\bf b}_{\alpha}^\c$ at positions
$\r_\alpha=(x_\alpha,0)$.
Crack dislocations interact with the external stress $\sigmab_{ij}^\ext$ and the
disorder made up of impurities and frozen dislocations.  For the further
discussion, it turns out to be convenient to introduce a 2-dimensional
dislocation density
        \begin{math}
        {\bf b}(\r)=\sum_{\alpha}{\bf b}_{\alpha}
        \delta(\r-\r_{\alpha})\,.
        \label{eq:2.2}
        \end{math}
        The actual distribution of the crack dislocation will be
        determined later from a minimum condition for the free energy
        for given external stresses and crystal imperfections.
It should be mentioned however, that the crack description by 
 dislocations is {\it  not} essential for the final results. We could also 
have used more traditional elasticity theory combined with the appropriate 
boundary conditions on the crack surface. In this sense also amorphous 
materials are included (but there are no frozen dislocations in this case).

The interaction between the external stress $\sigmab_{ij}^\ext$ and a
dislocation with Burgers vector ${\bf b}$ is given by the Peach-K\"ohler force
\cite{Landau}
        \begin{math}
        f_i=-\varepsilon_{il}^{\phantom{1}}b_m^{\phantom{1}}
        \sigmab_{lm}^\ext\,,
        \label{eq:2.3}
        \end{math}
where $\epsilon_{il}$ denotes the total antisymmetric unit tensor. With the
help of this relation one obtains for the total contribution of
$\sigmab_{ij}^\ext$ on the crack dislocation energy 
        \begin{equation}
        E^\ext_{\phantom{1}}=-\varepsilon_{xl}^{\phantom{1}}\sigmab_{lm}^\ext
        \sum\limits_{\alpha}x_{\alpha}^{\phantom{1}}b_{\alpha,m}^\c=
        -\sigmab_{ym}^\ext\inta{x} x \,b^\c_m(x)\,.
        \label{eq:2.4}
        \end{equation}
Thus, in mode I (II) only the $y$ ($x$) component of ${\bf b}^\c(x)$
contributes to $E^\ext$. Since $E^\ext$ is the only macroscopic term which
favors the existence of crack dislocations, it is clear that in mode I (II)
only those with ${\bf b}^\c({\bf r})$ parallel to the $y$-($x$-)axis will
occur. This will be used in the following \cite{C1}.

The stress field $\sigmab_{ij}$ generated by dislocations is related
to the Airy stress function $\chi({\bf r})$ by \cite{Kroener} 
        \begin{math}
        \sigmab_{ij}=\varepsilon_{ik}\varepsilon_{jl}\partial_k\partial_l
        \chi({\bf r})
        \label{eq:2.5}
        \end{math},
where $\chi$ is a solution of
        \begin{equation}
        (\bnabla^2)^2\chi({\bf r})=\Yb\varepsilon_{ji}\partial_jb_i({\bf r})\,.
        \label{eq:2.6}
        \end{equation}
Here $\Yb=4\mub(\lambdab+\mub)/(2\mub+\lambdab)$
denotes the Young modulus in two dimensions.
\begin{figure}
        \setlength{\unitlength}{1mm}
        \begin{picture}(85,43)(0,0)
        \put(0,0)
        {\epsfxsize = 3.9cm \epsfbox{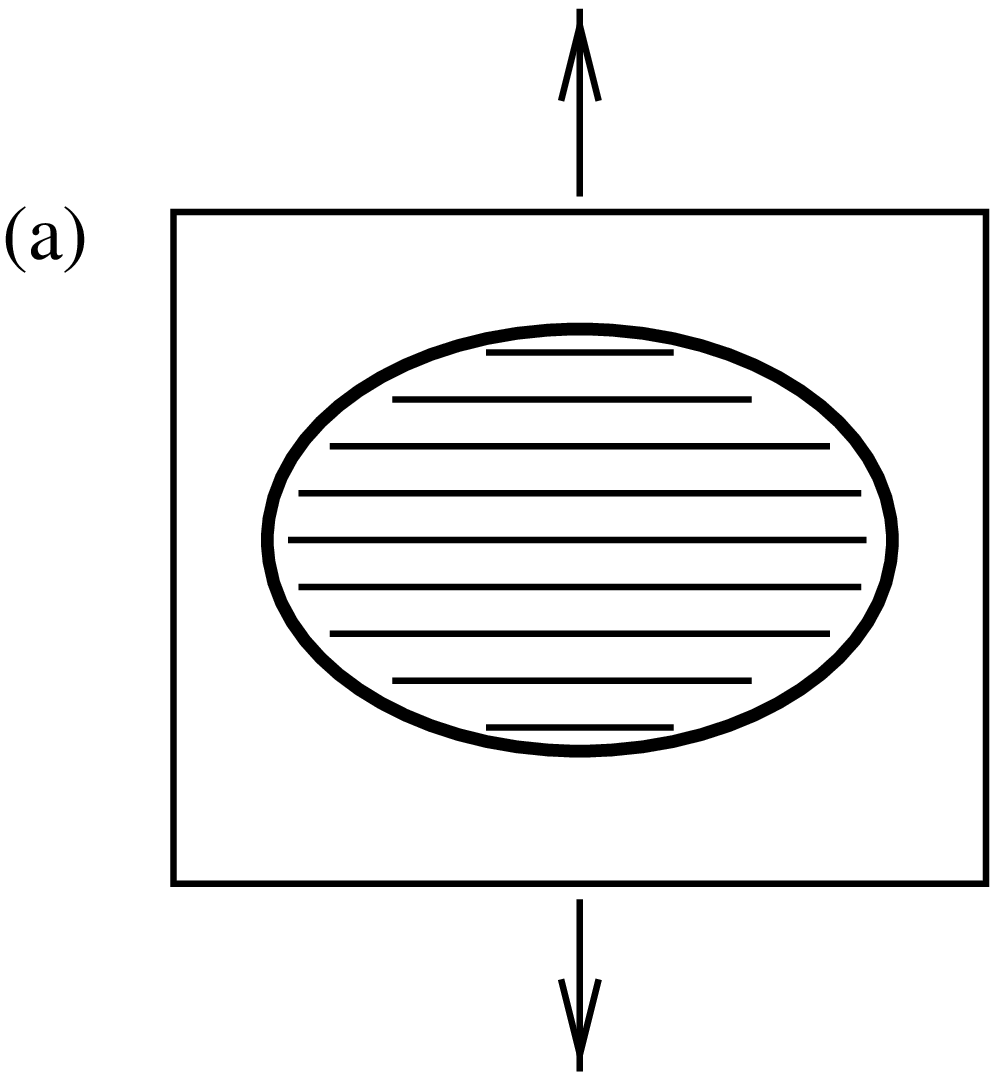}}
        \put(40,5)
        {\epsfxsize = 3.9cm \epsfbox{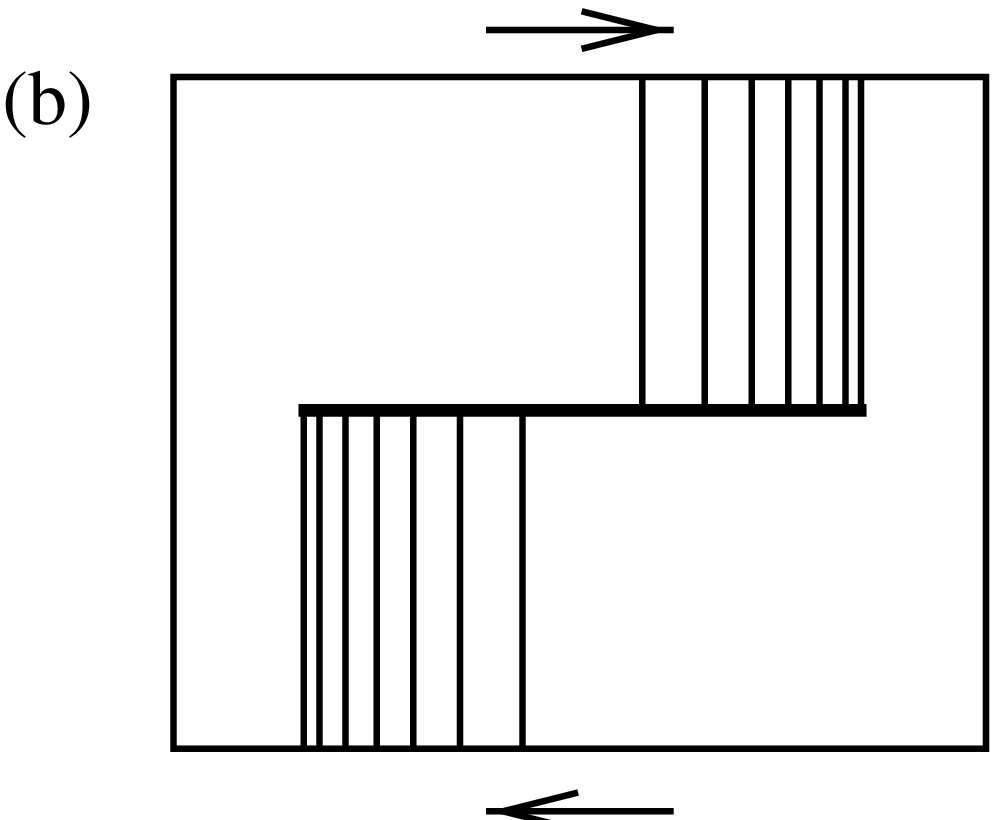}}
        \end{picture}
\caption{
\label{fig:1}
Lattice planes terminating in crack dislocations in mode I (a) and mode II
(b) cracks.  The dislocation vectors are perpendicular to the lattice planes.
The arrows denote the direction of the external forces $f^\ext$. Changing
$f^\ext$ crack dislocations in mode I and II climb and glide, respectively. 
}
\end{figure}
In an infinite system the solution of (\ref{eq:2.6}) is given by \cite{Landau}
        \begin{equation}
        \chi(\r)=\Yb \intt{r'} g(\r-\r') \epsilon_{ij} 
        \partial'_i b_j(\r')\,.
        \label{eq:2.7}
        \end{equation}
Here
        \begin{math}
        g(\r)=\r^2 (\log |\r| +C)/(8\pi)
        \label{eq:2.8}
        \end{math},
$C$ is an arbitrary constant.
The elastic energy of the dislocations is then given by
        \begin{eqnarray}
        E^\c &=& \frac12 \intt{r} \sigmab_{ij} u_{ij}
        \label{eq:2.9}
        \\
        &=& -\frac{\Yb}{2} \intt{r}\! \intt{r'} \epsilon_{ij}  \epsilon_{kl} 
        b_j(\r) b_l(\r') \partial_i \partial_k g(\r-\r')
	\nonumber
        \end{eqnarray}
where we used the relation
        \begin{equation}
        u_{ik}=\frac1{2\mub} \sigmab_{ik} 
        -\frac{\lambdab}{4\mub ( \lambdab+\mub)} \delta_{ik} \sigmab_{ll}.
        \label{eq:2.10}
        \end{equation}
The elastic
energy of the crack dislocations is then given by 
        \begin{equation}
        E^\c = -\frac{\Yb}{8\pi} 
        \inta{x} \inta{x'}  b^\c(x) b^\c(x') \log 
        \left|\frac{x-x'}{a_0}\right| \,.
        \label{eq:2.12}
        \end{equation}
$b^\c(x)$ denotes the crack dislocation density along the $x$-axis
and  $a_0$ denotes a microscopic cut-off (of the order of the lattice
spacing and in general different for mode I and mode II).  
Eq. (\ref{eq:2.12}) is valid both for mode I and mode II cracks.
Note that $E^\c$ is always positive.  

So far we did not consider the core contributions of the dislocations.
In the present context this is replaced by the 2-dimensional energy 
density $\gammab(x)=\gammab_0+\gammab_1(x)$
of the crack  surface
        \begin{equation}
        E^\sur = 2 \inta{x} \gammab(x)
        =4 \gammab_0 a + E^\sur_1(a)
        \label{eq:2.13}
        \end{equation}
$\gammab_1(x)$
reflects the randomness in the strength of the bonds
broken. For simplicity we assume Gaussian 
disorder with $\dav{\gammab_1(x)}=0 $ and 
       \begin{equation}
       \dav{\gammab_1(x)\gammab_1(x^{\prime})}=
       \dgamma^2 a_0\delta_{a_0}(x-x^{\prime})\,,
        \label{eq:davsur}
       \end{equation}
$\dav{\dots}$ denotes the average over the disorder and  $\delta_{a_0}(x)$ a
delta function of width $a_0$. 
In general the correlation length $a_0$ of the
disorder appearing in (\ref{eq:davsur}) is different from the cut-off appearing
in (\ref{eq:2.12}). Similarly, further correlation lengths could be introduced
for the distribution of impurities and frozen dislocations to be considered
below. To avoid a too clumsy notation we will however use everywhere the length
$a_0$ as a small scale cut-off but keeping in mind this complication.
Depending on the type of material under consideration, $a_0$ may vary between
the size of an atom in crystalline and the size of a grain in granular
materials, respectively.  The precise value of $a_0$ will be of course
important if comparison with experiments is made.  Then $\dav{E^\sur_1(a)}=0$
and we find for the variance of $E^\sur_1(a)$ \begin{equation}
        \dav{\left[E^\sur_1(a)-E^\sur_1(a')\right]^2 }=\Dsur|a-a'|
        \label{eq:2.14nn}
       \end{equation}
where $\Dsur=8 \dgamma^2 a_0$. Clearly Eq. (\ref{eq:davsur}) is only valid for
$|a-a'| \gtrsim a_0$.

In the following we add the contributions $E^\dis=E^\fd+E^\imp$ from randomly
frozen dislocations and impurities to the energy.  $E^\dis$ is given by
$\intt{r} \sigmab^\dis_{ij} u^\c_{ij}$ where $\sigmab^\dis_{ij}$ denotes the
stress generated by the disorder and $u^\c_{ij}$ the strain field
generated from cracks, respectively.

Using Eq. (\ref{eq:2.9})
$E^\dis$ can be written in the form
        \begin{equation}
        E^\dis = -\frac{\Yb}{4\pi} \inta{x} b^\c(x) V(x)\,,
        \label{eq:2.99}        
        \end{equation}
       \begin{equation}
         V(x)=V^\fd(x) + V^\imp(x)
        \label{eq:2.14}
       \end{equation}
{}From (\ref{eq:2.6}), (\ref{eq:2.7})
and  (\ref{eq:2.10}) one obtains for the
potential created from dislocations
        \begin{equation}
        V^\fd(x)= 4\pi \intt{r'} \epsilon_{ij} 
        \big(\partial_k \partial_i g(\r-\r')\big)_{y=0} b^\fd_j(\r')
        \label{eq:2.15}
        \end{equation}
where $k=x,y$ for mode I, II cracks, respectively.
The frozen dislocations are assumed to have both random positions $\r_\alpha$
and directions of their Burgers vectors such that
        \begin{math}
        \dav{\b^\fd(\r)} = 0
        \end{math}
and
        \begin{equation}
        \dav{b^\fd_i(\r) b^\fd_j(\r')} = 
        \bfd^2 c_\fd \delta_{a_0}(\r-\r') \delta_{ij}
        \label{eq:2.16}
        \end{equation}
Here $c_\fd$ and $\bfd$ denote the concentration and the strength of the
dislocation. 

Impurities (or more macroscopic inclusions) also generate a long-range elastic
displacement field $\u^\imp(\r)$.  Repeating the calculation of Eshelby
\cite{Es} for $d=2$ dimensions one finds for the strain tensor of an impurity
located at the origin
        \begin{equation}
        u_{ij}^\imp(\r) = \frac{\Omega}{2\pi} 
        \frac{\lambdab+\mub}{2\mub+\lambdab} \partial_i \partial_j \log|\r|
        \label{eq:2.17}
        \end{equation}
Here $\Omega$ denotes the 2-dimensional volume change due to the impurity which
can be of either sign.  The interaction energy between the crack dislocations
and the impurities of density $\cb_\imp(\r)$ takes the form
        \begin{equation}
        E^\imp
        =\frac12 \Omega  \intt{r} \sigmab_{ii}^\c(\r) \cb_\imp(\r)
        \label{eq:2.19}
        \end{equation}
Here we used (\ref{eq:2.17}) and 
        \begin{equation}
        \cb_\imp(\r)=\sum_\alpha \delta_{a_0} (\r-\r_\alpha) - \cb_\imp
        \label{eq:2.18}
        \end{equation}
where the summation is over all impurity sites $\r_\alpha$ and $\cb_\imp$ denotes
the impurity concentration.
With (\ref{eq:2.6})--(\ref{eq:2.10}) we find
        \begin{equation}
        V^\imp(x)= \Omega \intt{r'}  c(\r') \left(\partial_k
        \log \left| \frac{\r-\r'}{a_0} \right|  \right)_{y=0}
        \label{eq:2.20}
        \end{equation}
again $k=x,y$ for mode I, II cracks, respectively.

The total energy
        \begin{math}
        E=E^\ext+E^\c+E^\sur+E^\dis
        \label{eq:2.21}
        \end{math}
is   a functional of the crack dislocation density $b^\c(x)$.
Differentiating the saddle point equation
        \begin{math}
        {\delta E}/{\delta b^\c(x)} = 0
        \label{eq:2.22}
        \end{math}
with respect to $x$, we find
        \begin{equation}
        \frac{4\pi}{\Yb} \sigmab^\ext + \inta{x'} b^\c(x') \frac1{x-x'} +
        V'(x)=0
        \label{eq:2.23}
        \end{equation}
Eq. (\ref{eq:2.23}) has the solution \cite{Tricomi}
        \begin{eqnarray}
        b^\c(x) &= &\inta{x'} f(x,x';a) 
                (\frac{4\pi}{\Yb} \sigmab^\ext + V'(x') )
        \nonumber\\
        &=&b^\c_0(x)+b^\c_1(x)
        \label{eq:2.24}
        \end{eqnarray}
where
        \begin{equation}
        f(x,x';a)=-\frac{1}{\pi^2} 
                \left(\frac{a^2-x'^2}{a^2-x^2}\right)^{1/2}\frac{1}{x'-x}
        \label{eq:2.25}
        \end{equation}
The total energy as a function of the crack length follows with the help
of (\ref{eq:2.23})
        \begin{eqnarray}
        E(a)&=&\frac{\Yb}{8\pi} \inta{x} \inta{x'} b^\c(x) b^\c(x') 
        \log \left|\frac{x-x'}{a_0}\right| 
        \nonumber\\
        &&+2 \inta{x} \gammab(x)
        \label{eq:2.26}
        \end{eqnarray}
where $b^\c(x)$ given by (\ref{eq:2.24}).

For vanishing disorder $b^\c(x) \rightarrow  b^\c_0(x,a)$ for which
we obtain from (\ref{eq:2.24}), (\ref{eq:2.25}) and (\ref{eq:A.1})
        \begin{equation}
        b^\c_0(x,a)=\frac{4 \sigmab^\ext }{\Yb} \frac{x}{(a^2-x^2)^{1/2}}
        \label{eq:2.27} 
        \end{equation}
which yields in mode I an elliptic crack of maximal height 
$2 \sigmab^\ext a/\Yb$.

As follows from (\ref{eq:2.24}) the total energy (\ref{eq:2.26}) can be
divided into  contributions $E_n$ which are proportional to
$(\sigmab_{\ext})^{2-n}$ with $n=0,1,2$, respectively.  Here $E_1$ and
$E_2$ depend on the disorder. The disorder-independent contributions to the
energy are given by the Griffith expression 
 	\begin{equation}
	E_0(a)=4 \gammab_0 a -\frac{\pi a^2 \sigmab_\ext^2}{\Yb}
	=4 \gammab_0 a \bigg(1-\frac{a}{2 a_c}\bigg)
	\label{eq:2.28}
 	\end{equation}
which shows a maximum at $a=a_c=2 \gammab_0 \Yb/(\pi{\sigmab_{\ext}}^2) $ 
corresponding to an energy barrier $E_0(a_c)=2\gammab_0 a_c$.

The contributions $E_1$ and $E_2$ depend on the frozen
disorder and can be characterized by their mean value and variance.
$E_1$ can be rewritten using partial integration and (\ref{eq:A.1})
and (\ref{eq:A.2}) as
        \begin{eqnarray}
        E_1 & = & \frac{\Yb}{4\pi}
        \intinta{x}{x'} b_0^{\c}(x)b_1^{\c}(x^{\prime})
        \log{\left|\frac{x-x'}{a_0}\right|}
        \nonumber\\
        &   
        =& -\frac{\Yb}{4\pi} \inta{x} V(x) b_0^{\c}(x)
        \label{eq:2.29}
        \end{eqnarray}

For impurities we obtain from (\ref{eq:2.18}) and (\ref{eq:2.20})
        \begin{equation}
        \dav{V^{\imp}(x)V^{\imp}(x^{\prime})}=\Omega^2\cb_\imp \pi
        \log{\frac{R}{|x-x'|}}
        \label{eq:2.30}
        \end{equation}
where $R$ is a cut-off of the order of the system size which has to be send 
to infinity.  This gives with
(\ref{eq:A.1})--(\ref{eq:A.2}) for the impurity contribution $E_1^{\imp}$ to
the variance of $E_1$
        \begin{equation}
        \dav{\left[E_1^\cimp(a)-E_1^\cimp(a')\right]^2}
        =\Dimp|a^2-a'^2|
        \label{eq:2.31a}
        \end{equation}
Here $\Dimp = ({\pi}/{2}) \cb_\imp (\Omega \sigmab_{\ext})^2 = \cb_\imp \Omega^2\gammab_0 \Yb / a_c $. 

For frozen dislocations we get from (\ref{eq:2.15}) and (\ref{eq:2.16})
        \begin{equation}
        \dav{V^{\fd}(x)V^{\fd}(x^{\prime})}=\frac{6}{\pi}
        c_{\fd}\bfd^2\left(\frac{(\pi R)^2}{3}-(x-x^{\prime})^2\right)
        \label{eq:3.33}
        \end{equation}
which gives with (\ref{eq:2.29}) for the dislocation contribution $E_1^\fd$ to the
variance of $E_1$
        \begin{equation}
        \dav{\left[E_1^{\cfd}(a))-E_1^{\cfd}(a')\right]^2}
        =\Dfd(a^2-a'^2)^2 
        \label{eq:2.34a}
        \end{equation}
with
$\Dfd = ({3}/{\pi}) c_\fd  (\bfd \sigmab^\ext)^2 = 6 c_\fd \bfd^2
\gammab_0 \Yb / ( \pi^2 a_c) $.
It is easy to see from  (\ref{eq:2.99}) and the condition that the is crack 
closed, i.e. $\inta{x} b^\c(x)=0$, that there dependence on $R$ vanishes in 
(\ref{eq:2.31a}), (\ref{eq:2.34a}).
These equations are clearly valid only for $|a-a'|$ larger than $a_0$.
For $|a-a'|$ smaller than the mean distance between the impurities
or dislocations, respectively, the statistics of $E_1(a)$ is not longer Gaussian,
but Eqs. (\ref{eq:2.31a}), (\ref{eq:2.34a}) still give the correct order 
of magnitude of the fluctuations of $E_1(a)$.

\medskip

The mean values as well as the fluctuations of $E_2$ are proportional to
$\cb_\imp$ and $c_\fd$ and hence small if the disorder is weak as we will
assume in the following. 
Then
the average energy of the crack is given by the Griffith expression $E_0(a)$.
The energy barrier $E_0(a_c)=2\gammab_0 a_c$ is typically large and cannot
be overcome by thermal fluctuations.
Indeed, for  crystalline  solids with\cite{Griff} $\gammab_0\lesssim\Yb a_0$ 
one finds 
	\begin{equation}
 	E_0(a_c)\lesssim \kb T_m(\Yb/\sigmab)^2
	\,,\qquad
	T_m=\gammab_0a_0/\kb
	\end{equation}
where $T_m$ is a characteristic temperature comparable to, but 
typically bigger than, the solid's melting temperature\cite{Feng}.
For relatively large strain 
$\Yb/\sigmab$ is of the order $10$ such that the nucleation rate for a 
supercritical crack is of the order $\omega_0\exp(-100T_m/T)$!
 $\omega_0$ is a microscopic attempt frequency of the order $10^{13}\mbox{s}^{-1}$.
In the further discussion we will therefore mostly 
neglect thermal fluctuations.

Let us denote the probability that a crack of length $a$ has a 
negative energy by $W_{E<0}(a)$. The smallest  crack one can think of has 
a length of the order $a_0$. Thus a crack can  only appear
if   this smallest crack has a negative energy, $E(a_0)<0$. This occurs with  
the probability $W_{E<0}(a_0)$. (On this smallest 
scale even thermal fluctuations may help to create a crack as we will see 
below). The crack can 
then only propagate further, if for a given disorder 
configuration the force on the crack tip
$f(a)=-\partial E/\partial a $ is positive for {\em all} $a\ge a_0$ (we 
neglect effects of inertia). 

\begin{figure}
        \setlength{\unitlength}{1mm}
        \begin{picture}(85,55)(0,0)
        \put(10,0)
        {\epsfxsize = 6cm \epsfbox{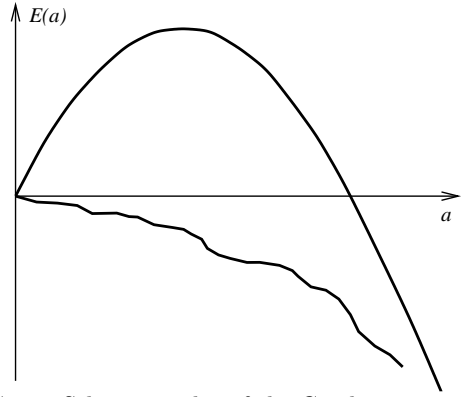}}
        \end{picture}
\caption{\label{fig:2}
Schematic plot of the Crack energy as a function of $a$ in the 
pure and random system.
}
\end{figure}

Because of its mathematical simplicity we consider here first a {\it necessary}
condition to be fulfilled which is $E(a)<0$ for {\it all} $a$. Clearly, 
if  $E(a)>0$ 
an (essentially) macroscopic energy barrier exist and the crack 
cannot propagate.
 The probability $W_{E<0}(a)$ 
is given by
        \begin{equation}
        W_{E<0}(a)=\frac1{\sqrt{2\pi}} 
        \int_{-\infty}^{-\phi(a)} \!{\rm d}x\,
        e^{-x^2/2}
        \label{eq229}
        \end{equation}
with
         \begin{equation}
                \phi(a)=\frac{4 \gammab_0 a [1-a/(2a_c)]}
                {(\Dsur a+\Dimp a^2+\Dfd a^4)^{1/2}}
	\label{phia}
         \end{equation}
Here we used the fact that $E_1^\sur+E_1^\cimp+E_1^\cfd$ is Gaussian
distributed with variance $\Dsur a+\Dimp a^2+\Dfd a^4$. Note that this
expression for the variance makes sense only for $a$ much larger than
$a_0$.  Among the cracks of various lengths $a$ there is one length
$\tilde{a}_c$ for which the probability ${W}_{E<0}$ is minimal.  This minimum
appears at the maximum of $\phi(a)$. One finds $\tilde{a}_c$ given by the
solution of
        \begin{equation}
        0= (1-\frac{3a}{2 a_c}) \Dsur  - \frac{a^2}{a_c} \Dimp -2 a^3 \Dfd
        \label{eq230}  
        \end{equation}

It is tempting to consider $W_{E<0}(\tilde{a}_c)\equiv\tilde{W}_{E<0}$
 (compare (\ref{eq229}), (\ref{eq230}))  as 
the probability for the
occurrence of the crack. This conclusion will be elucidated further below.  It
is instructive to consider the three different sources of disorder separately:

(i) If only the surface energy is random, then $\phi_\sur(a)$
vanishes at $a=0$ and $a=2 a_c$
and has an maximum at $\tilde{a}_c=(2/3) a_c$.
For weak disorder $\phi^\sur (\tilde{a}_c) \equiv \tilde{\phi}^\sur$
is large and since
for $\phi(a)\gg 1$, $W_{E<0}(a)\approx
-\exp(-\phi^2/2)/(\sqrt{2\pi} \phi)$ we get
        \begin{equation}
        \tilde{W}^\sur_{E<0}\approx
        \frac1{\sqrt{2\pi} \phi^\sur(\tilde{a}_c)} 
        \exp \bigg({-\frac{16\gammab_0^2 a_c}{27 \dgamma^2 a_0}}\bigg)
        \label{eq:2.47}
        \end{equation}
(ii) In the case of randomly distributed impurities  $\phi_\imp(a)$ 
has a maximum for vanishing $a$.  
For weak disorder and $a_c \gg a_0$ we get 
        \begin{equation} 
        \tilde{W}^\imp_{E<0}\approx 
        \frac1{\sqrt{2\pi} \phi^\imp(\tilde{a}_c)}
        \exp\bigg( {-\frac{8 \gammab_0 a_c}{\Omega^2 \cb_\imp \Yb}}\bigg)
        \label{eq:2.50}
        \end{equation}
(iii) Finally, for frozen dislocations
        \begin{math}
        \phi_\fd(a)=4 \gammab_0 (1/a-1/(2a_c))/\sqrt{\Dfd}.
        \end{math}
Hence $\phi_\fd(a)$ takes its maximum again at $a\rightarrow 0$ where $\phi(a)$ 
diverges. 
In this case $a$ 
has to be replaced by the minimal lengthscale $a_0$.
This gives 
        \begin{equation} 
        \tilde{W}^\fd_{E<0}\approx
         \frac1{\sqrt{2\pi} \phi^\fd(\tilde{a}_c)}
         \exp\bigg( {-\frac{4 \pi^2 \gammab_0 a_c}{3 \bfd^2 c_\cfd a_0^2  \Yb}}\bigg).
	\label{Wfd}
        \end{equation}

\medskip

It may be worthwhile to mention, that thermal fluctuations could be at 
least partially incorporated into the present treatment by relaxing the 
condition $E(a)<0$ to  $E(a)<\kb T\log(\omega_0 t)$ 
since  barriers of order $E$ are overcome on 
time scales of the order  $t\approx \omega_0\exp(E/T)$. 
This replacement 
changes the numerator of  $\phi(a)$ in eq. (\ref{phia}) from
$4 \gammab_0 a (1-a/(2a_c))$ to $4 \gammab_0 a (1-a/(2a_c))-\kb T \log(\omega_0 t)$.
Since barriers on scales $a \lesssim a_T$ with
	\begin{equation}
	a_T \approx a_0 \frac{T}{4 T_m} \log(\omega_0 t)
	\end{equation}
disappear, $a_0$ has to be replaced by $\max (a_0,a_T)$ in the expressions for
$\tilde{W}_{E<0}$ (we still assume $a_c \gg a_T$). This replacement effects
essentially only the result for the frozen dislocations, (\ref{Wfd}).
We emphasize however, that in general the three types of disorder will 
work in parallel.

\medskip

Next we consider the probability to fulfill the {\it sufficient}
condition $\partial E/\partial a <0$ for $0<a<\infty$, i.e.
that the force $f(a)$ acting on the crack tip is always
positive and the crack can propagate forever. We decompose $f$ into 
a deterministic and a stochastic contribution:. 
        \begin{equation} 
        f(a)=-\frac{\partial E(a)}{\partial a}
        =f_0(a)+f_1(a)
        \label{eq:2.51}
        \end{equation}
        \begin{equation} 
        f_0(a)=4 \gammab_0 \bigg(1-\frac{a}{a_c}\bigg)
        \label{eq:2.52}
        \end{equation}
        \begin{equation}
        f_1(a)=f_1^\sur(a)+f_1^\imp(a)+f_1^\fd(a)
        \end{equation}
Since $f_1(a)=-\partial E/\partial a$ is Gaussian distributed, the joint 
probability distribution
of $f_1(a)$ for $0<a<\infty$ is also Gaussian but in general {\it non-local}.
Its form can be reconstructed from the second moments
        \begin{eqnarray}
        \dav{f_1^\sur(a)f_1^\sur(a')} &=& \Dsur \delta_{a_0}(a-a')
        \nonumber \\
        \dav{f_1^\imp(a)f_1^\imp(a')} &=& 2 a \Dimp \delta_{a_0}(a-a')
        \nonumber \\
        \dav{f_1^\fd (a)f_1^\fd (a')} &=& 4 a a'\Dfd 
        \nonumber
        \end{eqnarray}
        In the case of random surface tension and randomly distributed
        impurities correlations are local and the joint probability
        distribution of the $f$'s factorizes. In this case the
        probability that the force  $f(a)$ on the tip of a crack of length 
$a$ is positive is given by
        \begin{equation} 
        W_{f>0}(a)=
        \int_{-\infty}^{-f_0(a)/\dav{f_1^2(a)}^{1/2}} \! \mbox{d} x\,
        \frac1{\sqrt{2\pi}} e^{-x^2/2}
        \label{eq:2.54}
        \end{equation}
The total probability $\tilde{W}_{f>0}$ for $f(a)>0$ for $0<a<\infty$ is given
by the product of all $W_{f>0}(a)$ factors. Here we take into account, that 
the forces are correlated over a distance of order $a_0$ such that we can 
decompose the crack in pieces of the order $a_0$. It is more convenient 
to consider the logarithm of $\tilde{W}_{f>0}$
        \begin{equation} 
        \log \tilde{W}_{f>0}=
        \sum_a \log 
        W_{f>0}(a)
        \label{eq:2.55}
        \end{equation}
The sum over $a=n a_0$ is here meant over integer numbers $n$. Since
for $a \gg a_c$ the integrals (\ref{eq:2.54}) are essential equal to unity,
it is sufficient to restrict the summation in (\ref{eq:2.55}) to the region
$0<a \lesssim a_c$. Moreover, the sum is dominated by the smallest values of
$W_{f>0}(a)$ for which we can replace the Gaussian integral
(\ref{eq:2.54}) by the approximate expression used above.  This gives
        \begin{equation} 
        \log \tilde{W}_{f>0} \approx
        -\int_0^{a_c} \frac{{\rm d} a }{2a_0} 
        \left(
        \frac{f^2_0(a)}{\dav{f_1^2(a)}}
        +\log \frac{\dav{f_1^2(a)}}{2 \pi f^2_0(a)}
        \right)
        \label{eq:2.60}
        \end{equation}
In the case of a random surface tension only we obtain
        \begin{equation} 
        \log \tilde{W}^\sur_{f>0} \approx
        -\frac{a_c}{a_0}\left( \frac{\gammab_0^2}{3\dgamma^2}   
        +1 + \frac{1}{2} \log \frac{\dgamma^2}{4 \pi \gammab_0^2}\right)
        \end{equation}
Thus $\tilde{W}^\sur_{f>0}$ is essentially of the same form as
$\tilde{W}_{E<0}$ apart from a replacement of the numerical factor in the
exponent ($\frac{16}{27}$ is replaced by $\frac1{3}$).
A similar calculation for the case of random impurities gives
        \begin{eqnarray} 
        \log \tilde{W}^\imp_{f>0} \approx
        -\frac{a_c}{a_0}\Bigg( \frac{4 \gammab_0 a_0}{\cb_\imp \Omega^2 \Yb}
        (\log \frac{a_c}{a_0} -\frac23 )
        \nonumber\\ 
        +1+\frac12 \log \frac{\cb_\imp \Omega^2 \Yb}{32 \pi a_0 \gammab_0} \Bigg)
        \end{eqnarray}

which is again the same result for the exponent as for $\tilde{W}_{E<0}$, apart
from the logarithmic factor which replaces 2.

The third case of randomly distributed dislocations is more involved.  Here we
should indeed take into account the long range correlations of the forces
$f_1^\fd$. This requires a more detailed mathematical investigation which we
leave for further studies. However our experience with the two other cases
makes it tempting to assume that $\tilde{W}^\fd_{E<0}$ gives essentially the
correct expression of the probability for the occurrence of a crack.

\section{Discussion and Conclusions}
It is interesting to remark that in all  cases considered above $\tilde{W}$ can
be written in the form of an Arrhenius law for thermal nucleation with an
effective temperature determined by the strength of the disorder:
        \begin{equation} 
        \tilde{W} \approx  \tilde{W_0}
        \exp \Big( - \frac{2 \gammab_0 a_c}{\kb T_{\rm eff}}\Big)
        \label{eq3.1}       
        \end{equation}
Here 
        \begin{equation}
        T_{\rm eff}^\sur=\frac{6 \dgamma^2 a_0}{\kb \gammab_0}
        \label{eq3.2}
        \end{equation}
for random surface tension, 
        \begin{equation}
        T_{\rm eff}^\imp=\frac{\Omega^2 \cb_\imp \Yb}{4\kb }
         \label{eq3.3}       
        \end{equation}
for randomly distributed impurities and 
        \begin{equation}
        T_{\rm eff}^\fd\approx \frac{3 a_0^2 \bfd^2 c_\fd \Yb}{2 \pi^2 \kb}
         \label{eq3.4}       
        \end{equation}
for frozen dislocations. The present calculation is not
accurate enough to determine the pre-exponential term $\tilde{W_0}$, which we
assume to be of the order one.  Relation (3.1) can be given a very simple
meaning: the probability that a crack of minimal length $a_0$ occurs is given
by $\exp ( - 2 \gammab_0 a_0/(\kb T_{\rm eff}))$. Here $2\gammab_0 a_0$
denotes the energy of such a crack. The total probability is then the
$(a_c/a_0)$-th power of this elementary probability.

Below we want to estimate $\tilde{W}$ for two different materials. We have to
keep in mind, that  our calculation was strictly two-dimensional. To compare
the results with real experiments on {\it thin plates} we have 
to consider their dependence on the width  $h$ of the plate.  
A necessary condition for the application of the two-dimensional theory 
is, that the critical crack length $a_c$ is {\it much larger} than $h$.
In the following we will assume, that this condition is always fulfilled.

Since $\gammab_0=\gamma_0h$,
$\Yb=Yh$ and $\sigmab_{\ext}=\sigma_{\ext} h $ are proportional to $h$, the
nucleation energy $2 \gammab_0 a_c=4 \gammab_0^2 \Yb/(\pi{\sigmab_{\ext}}^2)$
is also proportional to $h$. 

Estimating the $h$-dependence of $ T_{\rm eff}$ we have to make sure, that 
 all relevant length scales 
in the $xy$-plane (like $a_c$) are much larger than the $h$.
For $T_{\rm eff}^\imp$ and  $T_{\rm eff}^\fd$, which were determined by the 
small scale cut-off $a_0$, this leads to the severe restriction $ h < a_0$.
We note however, that  this cut-off will be in general larger 
than the lattice spacing. It is therefore appropriate to use in these 
two cases $h$ as the small scale cut-off. The behavior on even smaller scales 
is described by three-dimensional physics which we will discuss at the 
end of the section.

With  
$\dgamma^2 \approx \delta\gamma^2ha_0$, where $\delta\gamma$ denotes 
the fluctuation of the three-dimensional surface tension $\gamma_0$, 
$T_{\rm eff}^\sur $ does not depend on $h$.

Next, $\cb_\imp=c_\imp h$ is a two-dimensional density  proportional to $h$
(the three-dimensional density  $c_\imp$ is independent of $h$). $\Omega$ as a
two-dimensional cross section of the impurity should be replaced by
$\Omega^{3/2}/h$, i.e. we assume essentially spherical impurities.  Thus
        \begin{equation}
        T_{\rm eff}^\imp \rightarrow\frac{\Omega^3 c_\imp Y}{4\kb} \Big(1-\frac{h}{2a_c}\Big)
        \label{eq.imp.h}
        \end{equation} 
is essentially independent of $h  (\ll a_c)$.
   
For randomly distributed dislocations $c_\fd$ (as a line density) 
and $b_\fd$ are unchanged 
and hence   
        \begin{equation}
        T_{\rm eff}^\fd \rightarrow \frac{h^3 b_\fd^2 c_\fd Y}{4\pi^2\kb }
        \label{eq.fd.h}
        \end{equation} 
In (\ref{eq.imp.h}) and (\ref{eq.fd.h}) we replaced $a_0$ by $h$.

\bigskip

On scales smaller than $h$ the cracks are {\it three-dimensional}.  The results
of this paper can however easily be extended to penny cracks in $d$ dimensions.
The Griffith energy then takes the form 
        \begin{equation}
        E_0(a)=\gamma_0 a^{d-1} - Y^{-1} \sigma_\ext^2 a^d 
        \label{eq:ed}
        \end{equation} 
where we neglect here and in the following all numerical factors which in
general depend on the precise crack geometry.  $E_0(a)$ has  a maximum at $a_c
\sim\gamma_0Y/\sigma_\ext^2$, i.e. $a_c$ is essentially unchanged.

If the disorder is taken into account, additional
contributions to the energy appear.  A randomness in the surface tension leads
to a fluctuation of the order 
$\dav{(E^\sur_1(a))^2 } \approx \delta\gamma^2 (a a_0)^{d-1}$. 
Following the discussion below Eq. (\ref{eq229}), the minimum of  $W_{E<0}(a)$ 
follows again for $\tilde {a}_c \sim a_c$.
Rewriting the minimum $\tilde{W}_{E<0,d}$ of  $W_{E<0}(a)$ in the form
        \begin{equation}
        \tilde{W}_{d} \approx  \tilde{W}_{d,0}
        \exp \Big(- \frac{\gamma_0 a_c^{d-1}}{\kb T_{\rm eff}}\Big)
        \label{eq.tildeW.3d}
        \end{equation}
we obtain for the effective disorder temperature 
$T_{\rm eff}^\sur(d)\approx \delta\gamma^2 a_0^{d-1}/(\kb \gamma_0)$.

Randomly distributed impurities create an additional contribution $\sigma_\imp$
to the stress where 
        \begin{equation}
        |\sigma_\imp|= Y (\Omega^d c_\imp / a^d)^{1/2}
        \label{eq:2.32}
        \end{equation}
This expression can be understood as follows: An isolated impurity creates in a
volume $a^d$ an average stress \cite{Landau} of the order $Y \Omega^{d/2}/a^d$ .
With $c_\imp a^d$ the number of impurities in this volume (the average stress
created by the impurities is assumed to be already incorporated into $E_0(a)$)
the fluctuations of the stress created by the impurities is given 
by (\ref{eq:2.32}).
Thus we obtain $ \dav{(E_1^\cimp(a))^2}\approx \Omega^{d}Y\gamma_0 a^d c_\imp
/a_c$.  The minimal probability follows for $d>2$ (contrary to the
$2$-dimensional case) for $\tilde a_c\sim a_c$. The disorder temperature is now
of the order $T_{\rm eff}^\imp(d)=\Omega^d c_\imp Y/(4\kb )$.  Clearly,
for $a_c\approx h$,  $ T_{\rm eff}^\imp(d)$ agrees with the result for the
plate, as it should be.

Similarly, randomly distributed dislocation lines will give a fluctuation
contribution $\sigma_\fd$ to the stress of the order
        \begin{equation}
        |\sigma_\fd|= Y b_\fd c_\fd ^{1/2}
        \label{eq:2.33}
        \end{equation}
Indeed, a single dislocation (line) creates a stress of the order $Y b_0/a$.
With $c_\fd a^2$ for the number of dislocation lines in the volume $a^d$ which
are assumed to have random  orientations one obtains (\ref{eq:2.33}).  
The corresponding
fluctuation of the energy due to dislocations is therefore given $
\dav{(E_1^{\cfd}(a))^2} \approx a^{2d} \gamma_0 Y b_\fd^2 c_\fd /a_c$.  The
minimal probability follows as for $d=2$ from small scales $a\approx a_0$,
which results in  
        \begin{equation}
        T_{\rm eff}^\fd (d) \approx a_0^2 a_c^{d-2} Y b_\fd^2 c_\fd/\kb 
        \label{eq.fd.d}
        \end{equation}

Clearly, also the case of multiple disorder can be considered as it 
was done in (\ref{eq229}).

\begin{table}[t]
\begin{tabular}{llll}
         	&\hspace{2cm}         		&Glass\hspace{17mm}    		&SiC
\\
\hline
$Y$ 		&$[10^{9} \mbox{Pa}]$   	&70                   		&400
\\
$\gamma_0$ 	& $[\mbox{Jm}^{-2}]$    	&1.0         			&4.0 
\\
\hline
\hline
\multicolumn{4}{c}{random surface energy}\\
\hline
\multicolumn{4}{l}{weak disorder: $\delta\gamma/\gamma_0=0.1$, $a_0=5 \cdot 10^{-10}$m}
\\
$T_{\rm eff}^\sur (d=2)$&[K]      		&1087               		&4348 
\\
$A^\sur$ 	& $[\mbox{Pa}^2\mbox{m}^{-1}]$ 	&$\num{5.9}{30}$ 		&$\num{1.36}{32}$
\\
\hline
\multicolumn{4}{l}{strong disorder: $\delta\gamma/\gamma_0=0.3$, $a_0=10^{-6}$m}
\\
$T_{\rm eff}^\sur (d=2)$&[K]      		&$\num{3.9}{10}$               	&$\num{1.57}{11}$
\\
$A^\sur$ 	& $[\mbox{Pa}^2 \mbox{m}^{-1}]$ &$\num{1.65}{23}$              	&$\num{3.77}{24}$
\\
\hline
\hline
\multicolumn{4}{c}{random impurities}\\
\hline
\multicolumn{4}{l}{weak disorder: $\Omega=\num{2.5}{-19}$m${}^2$,$c=\num{8}{24}$m${}^{-3}$}
\\
$T_{\rm eff}^\imp (d=2)$&[K]     		&158.5                          &905.8
\\
$A^\imp$ 	&$ [\mbox{Pa}^2 \mbox{m}^{-1}]$ &$\num{4.07}{31}$ 		&$\num{6.5}{32}$
\\
\hline
\multicolumn{4}{l}{strong disorder: $\Omega=\enum{-15}$m${}^2$,$c=\enum{17}$m${}^{-3}$}
\\
$T_{\rm eff}^\imp (d=2)$&[K]     		&$\num{1.26}{5}$             	&$\num{7.24}{5}$ 
\\
$A^\imp$ 	&$ [\mbox{Pa}^2 \mbox{m}^{-1}]$ &$\num{5.09}{28}$ 		&$\num{4.15}{29}$
\\
\hline
\hline
\multicolumn{4}{c}{random frozen dislocations}\\
\hline
\multicolumn{4}{l}{weak disorder: $\bfd=\num{5}{-10}$m,$c_\fd=\enum{14}$m${}^{-2}$,$h=\enum{-3}$m}
\\
$T_{\rm eff}^\fd ({\rm plate})$ &[K]      	&                       	&$\num{1.83}{19}$ 
\\
$a_0=\num{5}{-9}$m\\
$A^\fd$ & $[\mbox{Pa}^2 \mbox{m}^{-1}]$ 	&                        	&$\num{1.71}{30}$
\\
\hline
\multicolumn{4}{l}{strong disorder: $\bfd\!=\num{5}{-10}$m,$c_\fd\!=\enum{16}$m${}^{-2}$,$h=\enum{-3}$m}
\\
$T_{\rm eff}^\fd ({\rm plate})$ &[K]      	&   		            	&$\num{1.83}{21}$ 
\\
$a_0=\num{5}{-10}$m\\
$A^\fd$ 	& $[\mbox{Pa}^2 \mbox{m}^{-1}]$ &                        	&$\num{1.71}{31}$
\\
\end{tabular}
\caption{
\label{tab}
Estimates of the effective temperatures $T_{\rm eff}$ for Glass and
SiC using (\ref{eq3.2}), (\ref{eq.imp.h}) and (\ref{eq.fd.h}).  The
corresponding factors $A$ in (\ref{eqas}) and (\ref{eqad}) are also given.  The
material constants are taken from \protect\ocite{Lawn}.}
\end{table}

Next we have to compare the different probabilities  in order to decide, which 
process dominates the formation of cracks in thin plates.
For systems with a random surface tension, and under the condition $a_c\gg h$, 
the minimal probability $\tilde{W}$ arises both in two and in three 
dimensions from cracks of length 
$\tilde{a}_c\sim a_c$. In this case the probability for crack  formation 
is given by (\ref{eq3.1}), (\ref{eq3.2}).

For randomly distributed impurities $\tilde{a}_c(d=2)\approx a_0\approx h$ in 
two dimensions and $\tilde{a}_c(d=3)\approx a_c$ in three dimensions. Thus 
crack formation is dominated here by two-dimensional cracks of minimal
length $a_0\approx h$. The corresponding probability is now given by 
(\ref{eq3.1}), (\ref{eq.imp.h}).

Finally, for randomly distributed dislocation, crack formation in both
two and three dimensions is controlled by the formation of small cracks of 
size $\tilde{a}_c\approx a_0$. Comparing the corresponding probabilities in 
two and in three dimension (for $d=2$ we have to use $a_0\approx h$) we find 
that crack formation is dominated by penny cracks of size $a_0$. Its 
probability is given by (\ref{eq.tildeW.3d}) and (\ref{eq.fd.d}) for $d=3$.

In a macroscopic sample of linear size $L$ regions of distance  
greater than $\tilde a_c$ can be considered to be essentially independent.
The total probability for crack formation is given by 
$\tilde W_d (L/\tilde a_c)^d$
where $d=2$ for surface and impurity disorder and $d=3$ for frozen 
dislocations. \cite{footnote}

Next we consider the probability for crack formation in two different 
materials, one is amorphous (glass) and one crystalline (SiC). The 
corresponding parameters are summarized in Table 1.
In the case of random surface tension and randomly distributed impurities 
we  express the probability for crack formation as a function of 
$h/\sigma_{(e)}^2$ since the probability can be written in the form 
        \begin{equation}
        \log(\tilde W/\tilde W_0)=
        -\frac{4}{\pi}\frac{\gamma_0^2 Y}{\kb  T_{\rm eff}}
        \frac{h}{\sigma_\ext^2}
        \equiv
        -A \frac{h}{\sigma_\ext^2}
        \label{eqas}
        \end{equation} 
For frozen dislocations we have
         \begin{equation}
        \log(\tilde W/\tilde W_0)\approx
        -\frac{\gamma_0^2}{a_0^3\bfd^2 c_\fd}\frac{a_0}{\sigma_\ext^2}
        \equiv 
        -A^\fd \frac{a_0}{\sigma_\ext^2}
        \label{eqad}
        \end{equation} 
The corresponding values for $A^\sur$, $A^\imp$ and $A^\fd$ are also given 
in Table \ref{tab}.

\medskip

A number of comments are in order: 

(i) The results so far are based on the
assumption of short range correlations of the disorder (surface energy,
impurities, frozen dislocations). Experimentally this may not be the most
important situation.  Long range correlations of the disorder described by a
power law decay on the r.h.s.  of (\ref{eq:davsur}), (\ref{eq:2.16}) and
(\ref{eq:2.19}) would in general lead to an increase of the probability for the
occurrence of a crack.

(ii) In the case of impurities and frozen dislocations the probability $\tilde{W}$
is dominated by the energetics on small length scales. Since the stress on the
crack tip diverges in the linear elasticity theory used throughout the paper,
non-linear effects may be of particular importance for small cracks.  This could
diminish the numerical coefficients in the exponent of $\tilde{W}$.

(iii) Defects were assumed everywhere to be frozen. In real fatigue 
experiments often
alternating stress  is applied which leads to an accumulation of
dislocations close to the crack tip, which makes crack propagation easier. This
mechanism could also help in a situation, were due to disorder fluctuations,
the  stress is considerably higher than in the average (or the surface energy
is lower). 

(iv) An interesting question is the relation of our results to crack 
propagation in quasicrystals \cite{Gumbsch}, which take an intermediate 
position between periodic and random media.

\medskip

We are very grateful to P. Gumbsch, V. Pokrovsky,  S. Scheidl and V.M. Vinokur 
for most valuable discussions.
This work is supported by the SFB 341 K\"oln-J\"ulich-Aachen.

\appendix
\section{Relevant Integrals}

        \begin{equation}
        I_1=\into{x'} (1-x'^2)^{1/2} \frac1{x'-x}= -\pi x
        \qquad {\rm for\ } |x|<1
        \label{eq:A.1}
        \end{equation}

        \begin{eqnarray}
        I_2&=&\into{x'} (1-x'^2)^{-1/2} \frac{x'}{x'-x}
        \nonumber\\
        &=&
        \left\{
                \begin{array}{ll}
                \pi & \mbox{ for } |x|<1 \\
                \pi -\pi |x| (x^2-1)^{-1/2} & \mbox{ for } |x|>1  \\
                \end{array}
        \right.
        \label{eq:A.2}
        \end{eqnarray}

\end{document}